# Inverse-Designed High-Q and Small-Mode-Volume Silicon Nitride Photonic Crystal Cavities for Second- and Third-Harmonic Generation

**M. Takiguchi[1,2], P. Heidt[2], X. Z. Lim[2], J. Zöllner[2], S. Yanagimoto[2], K. Nakayama[3], T. Aihara[4], M. Ono[1,2], H. Sumikura[1,2], and M. Notomi[1,2,5]**

*[1]NTT Nanophotonics Center, NTT, Inc., 3-1 Morinosato Wakamiya, Atsugi, Kanagawa 243–0198, Japan*

*[2]Basic Research Laboratories, NTT, Inc., 3-1 Morinosato Wakamiya, Atsugi, Kanagawa 243–0198, Japan*

*[3]International Center for Elementary Particle Physics, University of Tokyo, Bunkyo-ku, Tokyo 113-0033, Japan.*

*[4]Device Technology Laboratories, NTT, Inc., 3-1 Morinosato Wakamiya, Atsugi, Kanagawa 243–0198, Japan*

*[5]Department of Physics, Institute of Science Tokyo, Meguro-ku, Tokyo 152-8551, Japan*

*Author e-mail address: masato.takiguchi@ntt.com*

## Abstract

SiN photonic crystal (PhC) cavities are promising platforms for nonlinear and quantum photonics because of their wide transparency window, CMOS compatibility, and negligible two-photon absorption. However, realizing high-Q/V cavities remains challenging because of the relatively low refractive index of SiN. Here, we employ inverse design to optimize a two-dimensional SiN PhC cavity and experimentally demonstrate a quality factor of approximately 80,000, the highest reported for a near-stoichiometric SiN 2D PhC cavity. Furthermore, both second- and third-harmonic generation are observed from the same cavity, providing experimental evidence of strong optical confinement and large *Q/V*. Our results establish inverse-designed SiN PhC cavities as a promising platform for nonlinear photonics and future heterogeneous integrated photonic devices.

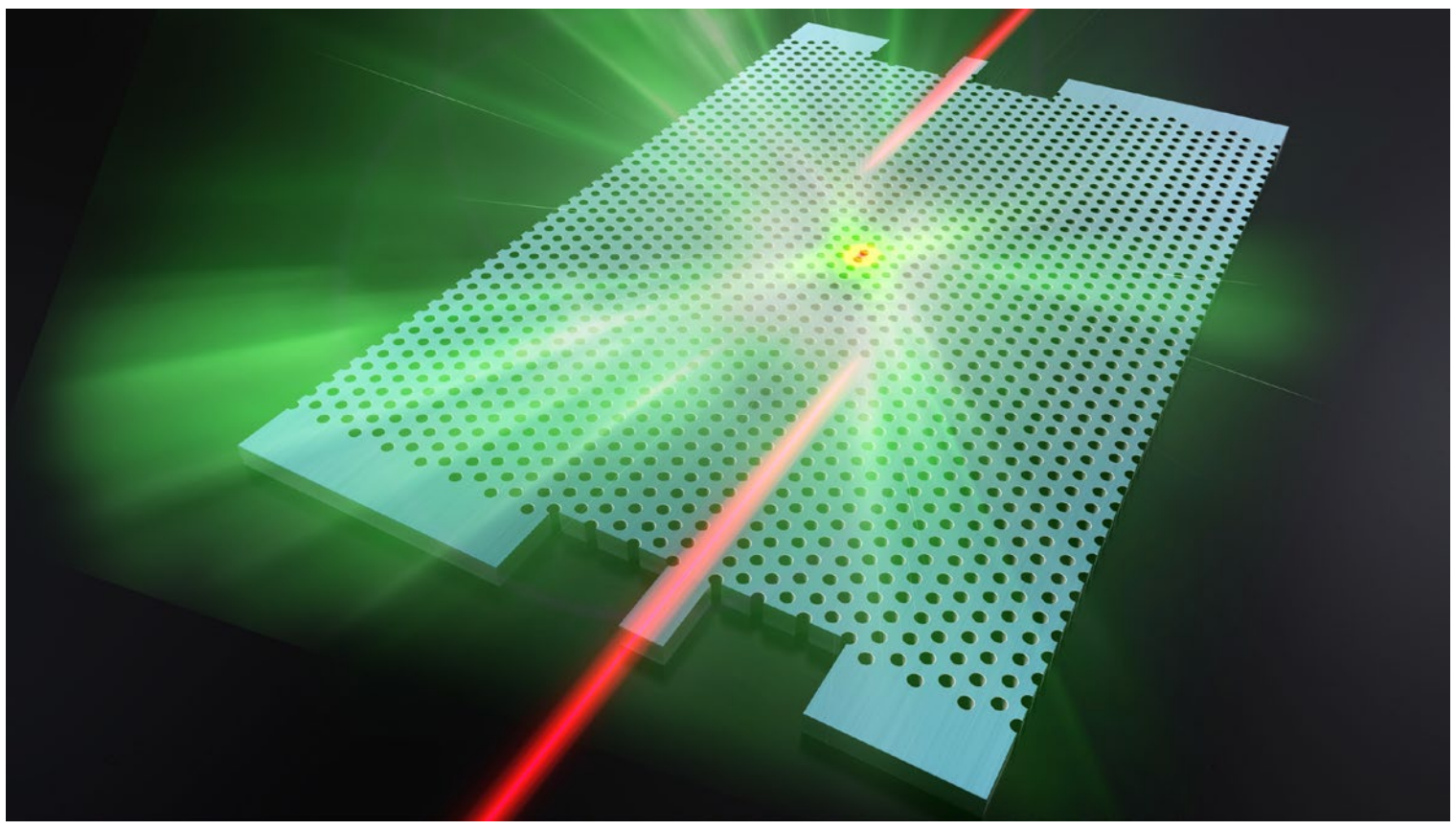

Fig. 1. :**Schematic of the SiN-based L6/5 photonic crystal cavity.**
Schematic of the SiN-based L6/5 photonic crystal cavity. The incident light is coupled into the cavity through the waveguide, where high-order harmonics are generated.

## Introduction

Recent advances in silicon nitride (SiN) photonics have established SiN as one of the most versatile platforms for integrated photonic devices, with widespread availability through commercial foundry services and applications ranging from fundamental research to practical technologies. Owing to its CMOS compatibility, low optical loss, and broad transparency window extending from the visible to the telecommunication wavelength range, SiN has attracted considerable attention for on-chip atomic spectroscopy[1,2,3], optical frequency combs[4,5], quantum light sources[6,7,8], and biosensing[9,10].

With a refractive index of approximately 2, SiN provides stronger optical confinement than low-index platforms such as silica-based planar lightwave circuits (PLCs), enabling more compact photonic devices. Furthermore, unlike silicon, SiN exhibits negligible two-photon absorption at telecommunication wavelengths, allowing significantly higher optical intensities to be introduced into waveguides. Consequently, by exploiting the third-order nonlinear susceptibility $\chi^{(3)}$ at high optical intensities, SiN has become a leading platform for Kerr nonlinear photonics[11,12,13], including optical frequency combs and dissipative Kerr solitons. In addition, because SiN remains transparent throughout the visible wavelength range, it is also an attractive platform for visible integrated photonics. Shorter operating wavelengths enable further miniaturization of photonic devices while facilitating heterogeneous integration with ultraviolet-to-near-infrared emitters and functional materials, including GaN[14], ZnO[15], diamond[16], rare-earth-ion-doped materials[17], organic emitters[18], particle emitters[19], organic dye molecules[20], quantum dots[21], and two-dimensional materials[22]. These applications require optical cavities with a large quality-factor-to-mode-volume ratio ($Q/V$) to maximize light–matter interactions. However, the relatively low refractive index of SiN compared with silicon weakens optical confinement, making it challenging to simultaneously achieve a high-$Q$ factor and a small mode volume.

Photonic crystal (PhC) cavities provide one of the most promising approaches for realizing extremely large $Q/V$ ratios. In particular, silicon enables a large refractive-index contrast between the cavity material and air, allowing two-dimensional PhC cavities to achieve exceptionally high $Q$ factors [23,24,25]. In contrast, near-stoichiometric $Si_3N_4$ has a relatively low refractive index of approximately 2.0, which limits the achievable Q factors of two-dimensional PhC cavities[26,27,28]. Only a few exceptions have been reported, including silicon-rich SiN two-dimensional PhC cavities[29,30] and SiN nanobeam cavities[31,32]. Conventional deterministic optimization methods improve cavity performance by introducing small perturbations to the air holes surrounding the cavity; however, this approach has been insufficient for realizing ultrahigh-*Q* SiN PhC cavities. Recently, inverse-design techniques[33,34,35,36], including machine learning[37,38] and optimization-based approaches, have emerged as powerful tools for photonic device design, enabling dramatic improvements in device performance across a wide range of applications. Such optimization techniques are particularly attractive because they enable the design of high-*Q/V* cavities even in resonators based on low-refractive-index materials, where such performance cannot be achieved using conventional deterministic design methods.

In this study, we optimized a two-dimensional SiN PhC cavity using an inverse-design approach and experimentally demonstrated a cavity with a high $Q/V$ ratio (Fig. 1). Furthermore, we observed both second-harmonic generation (SHG) and third-harmonic generation (THG) from the optimized cavity. Although SiN possesses a third-order nonlinear susceptibility, THG generally requires high optical intensities. In the present device, the strong optical-field enhancement provided by the inverse-designed high-$Q/V$ cavity enabled the observation of THG. In contrast, an effective surface second-order susceptibility, ${\chi_s}^{(2)}$, characterizes the nonlinear response at surfaces and interfaces where inversion symmetry is broken, making SHG generally difficult to observe. In our PhC cavity, however, the optical field is strongly localized near the surfaces of the air holes, allowing the surface ${\chi_s}^{(2)}$ response to be efficiently excited and SHG to be observed. The observation of these nonlinear optical processes provides compelling evidence of the large $Q/V$ ratio achieved by the inverse-designed cavity and highlights its potential for nonlinear photonics, wavelength conversion, and future heterogeneously integrated photonic devices.

## Results

### Inverse design

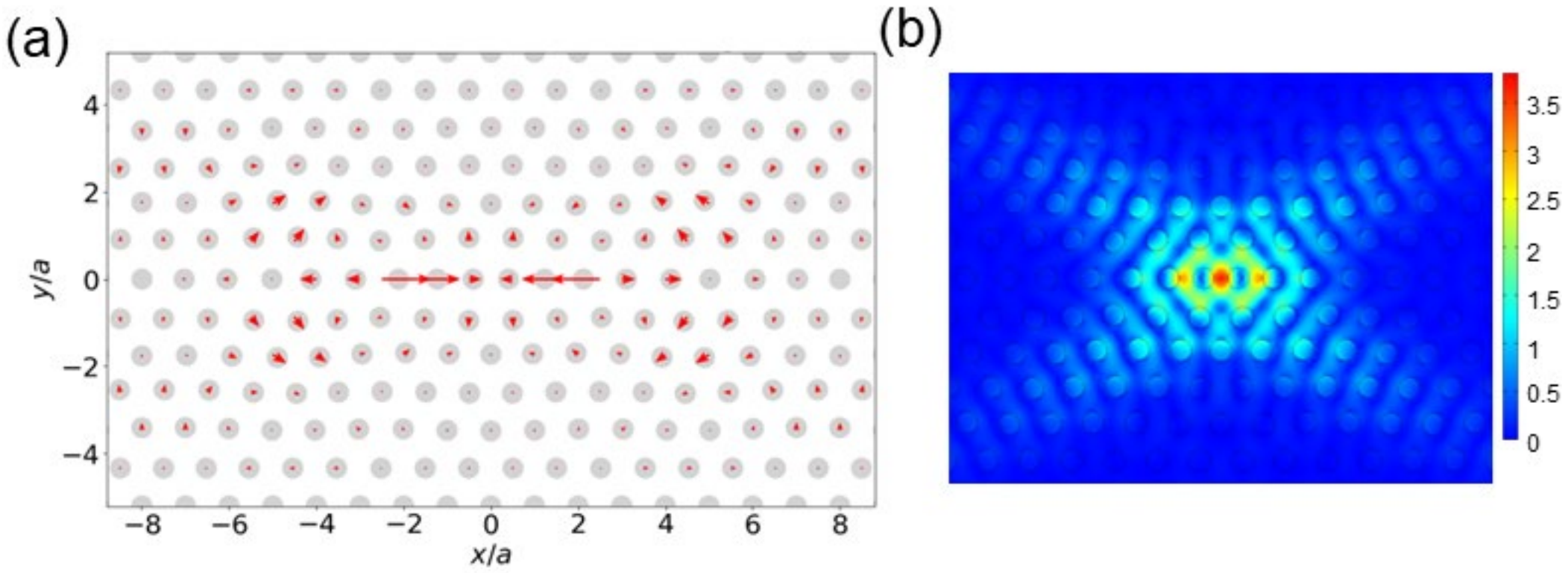

Fig. 2. : **Design and simulation of the inverse-designed L6/5 photonic crystal cavity.** (a) Inverse-designed L6/5 photonic crystal cavity. The arrows indicate the hole-position modulation vectors. For clarity, the arrow lengths are enlarged by a factor of three relative to the actual displacement in real space. (b) Electric-field distribution of the fundamental mode of the L6/5 cavity.

The relatively low refractive index of SiN results in a limited photonic bandgap, making it difficult to realize high-Q PhC cavities using conventional deterministic design methods. To overcome this limitation, we employed an inverse-design approach to optimize a SiN-based PhC cavity. The initial cavity geometry was based on an L6/5 cavity, which is expected to provide a larger $Q/V$ ratio than conventional cavity designs. In general, a cavity formed by removing three air holes from a photonic crystal lattice is referred to as an L3 cavity, whereas a cavity with four additional holes inserted into the defect region is known as an L4/3 cavity[39]. Such L(x+1)/Lx cavities are widely employed to achieve high $Q/V$ ratios. Structurally, the L4/3 and L6/5 cavities follow the same design; however, the L6/5 cavity was adopted in this work. Since inverse optimization is sensitive to the initial geometry, different starting structures may converge to different local optima. As shown in Fig. 2(a), the hole displacements are largest near the center of the cavity, suggesting that the inverse-design results are strongly influenced by the initial hole configuration in this region. In contrast, as x increases, the variation in the optimized hole displacements becomes much smaller, indicating that the influence of the initial configuration is likely to be limited. We also investigated an L8/7 cavity. Although it exhibited a slightly higher $Q$ factor, no significant improvement in the mode volume was observed.

The optimization targeted operation in the telecommunication wavelength range. The cavity geometry was optimized using the parallel simplex algorithm[40], also referred to as the parallel Nelder–Mead method, a derivative-free inverse-design method that can be readily implemented with conventional electromagnetic simulators without requiring gradient information or training data (see the Methods and Supplementary Note S1 for details). The optimization assumed an air-bridge cavity geometry to reflect the intended experimental device. The refractive index, slab thickness, and lattice constant were set to 2.0, 342 nm, and 686 nm, respectively. To reduce the computational cost, mirror symmetry was imposed with respect to the x-, y-, and z-planes, reducing the simulation domain to one-eighth of the full structure. The positions of 54 air holes surrounding the cavity were used as optimization parameters. Figure 2(a) shows the optimized hole displacements normalized by the lattice constant. Significant positional modulation is observed for many air holes surrounding the cavity center, demonstrating that the inverse-design algorithm exploits a large parameter space rather than modifying only the nearest neighboring holes. Figure 2(b) shows the electric-field distribution calculated by finite-element-method (FEM) simulations. The optimized fundamental cavity mode is located at 1551 nm with a simulated quality factor of $4.4 \times 10^5$ and a mode volume of approximately $0.96(\lambda/n)^3$, yielding an exceptionally large $Q/V$ ratio. We also confirmed that particle swarm optimization strategies produced comparable $Q/V$ values (see Supplementary Note S1). In addition to the fundamental mode, a higher-order cavity mode was found at 1519 nm with a simulated quality factor of approximately $2.7 \times 10^3$.

## Experimental characterization

Figure 3(a) shows a scanning electron microscope (SEM) image of the fabricated L6/5 PhC cavity (See the Supplementary Note S2 for additional details). The hole diameter, lattice constant, and slab thickness are 350 nm, 660 nm, and 350 nm, respectively. The cavity has an air-bridge structure and is integrated with input and output waveguides. As shown in the figure, mode converters are integrated at both ends of the input and output waveguides, which are connected to the chip facets through W8 PhC waveguides. This configuration is adopted because it is difficult to fabricate conventional narrow strip waveguides in an air-bridge structure (A suspended structure is required for the narrow waveguide[41]). The use of W8 PhC waveguides for optical routing is a well-established approach in InP-based photonic crystal devices[42]. In addition, separate input and output waveguides are coupled to the cavity from opposite sides, which facilitates the characterization of transmission through the cavity and allows the waveguide–cavity coupling to be readily controlled[43] . This configuration is also advantageous for experiments involving optical switching and related functionalities [44] [45] . Figure 3(b) shows the transmission spectrum of the cavity measured by sweeping the wavelength of a tunable laser (See the Supplementary Note S3). As shown in the figure, both the fundamental cavity mode (1462 nm) and second-order mode (1417 nm) are clearly observed. As discussed in the previous section, the wavelength spacing between adjacent cavity resonances is approximately 45 nm, which is in good agreement with the simulated results when the fabrication-induced deviations in the lattice constant and slab thickness are taken into account. The *Q*-factor of the second-order mode was also on the order of several thousand, in good agreement with the simulation results. Additional characterization of the PhC waveguides is provided in the Supplementary Note S4. In particular, an increase in the group index near the photonic crystal band edge, a characteristic feature of photonic crystal waveguides, was clearly observed. Figure 3(c) shows the transmission spectrum of the device exhibiting the highest quality factor among the fabricated samples. A Lorentzian fit to the resonance yields a quality factor of approximately $8.0 \times 10^4$. This quality factor significantly exceeds those reported for conventional stoichiometric SiN PhC cavities.

Next, we investigated the dependence of the cavity *Q*-factor on the relative position of the input/output waveguides. Figure 4(a) compares the measured and simulated quality factors for different cavity–waveguide coupling configurations, while the corresponding waveguide positions are illustrated in Fig. 4(b). Representative transmission spectra for the different coupling configurations are provided in Supplementary Note S4. The waveguides were placed at positions denoted as d4–d7 according to their distance from the cavity. As shown in Fig. 4(b), the d6 configuration corresponds to a modified geometry (d6-2), in which the cavity–waveguide coupling was intentionally reduced compared with the standard d6 position. In the simulations, the intrinsic cavity quality factor was assumed to be $4.4 \times 10^5$. The waveguide coupling quality factor ($Q_{\mathrm{coupling}}$) was first evaluated from the quality-factor degradation when a single waveguide was coupled to the cavity. The total quality factor for the two-waveguide configuration was then calculated and compared with the experimental results. As shown in Fig. 4(a), the measured quality factors exhibit excellent agreement with the simulations over the entire range of waveguide positions, confirming both the validity of the cavity design and the successful fabrication of the optimized SiN PhC cavities.

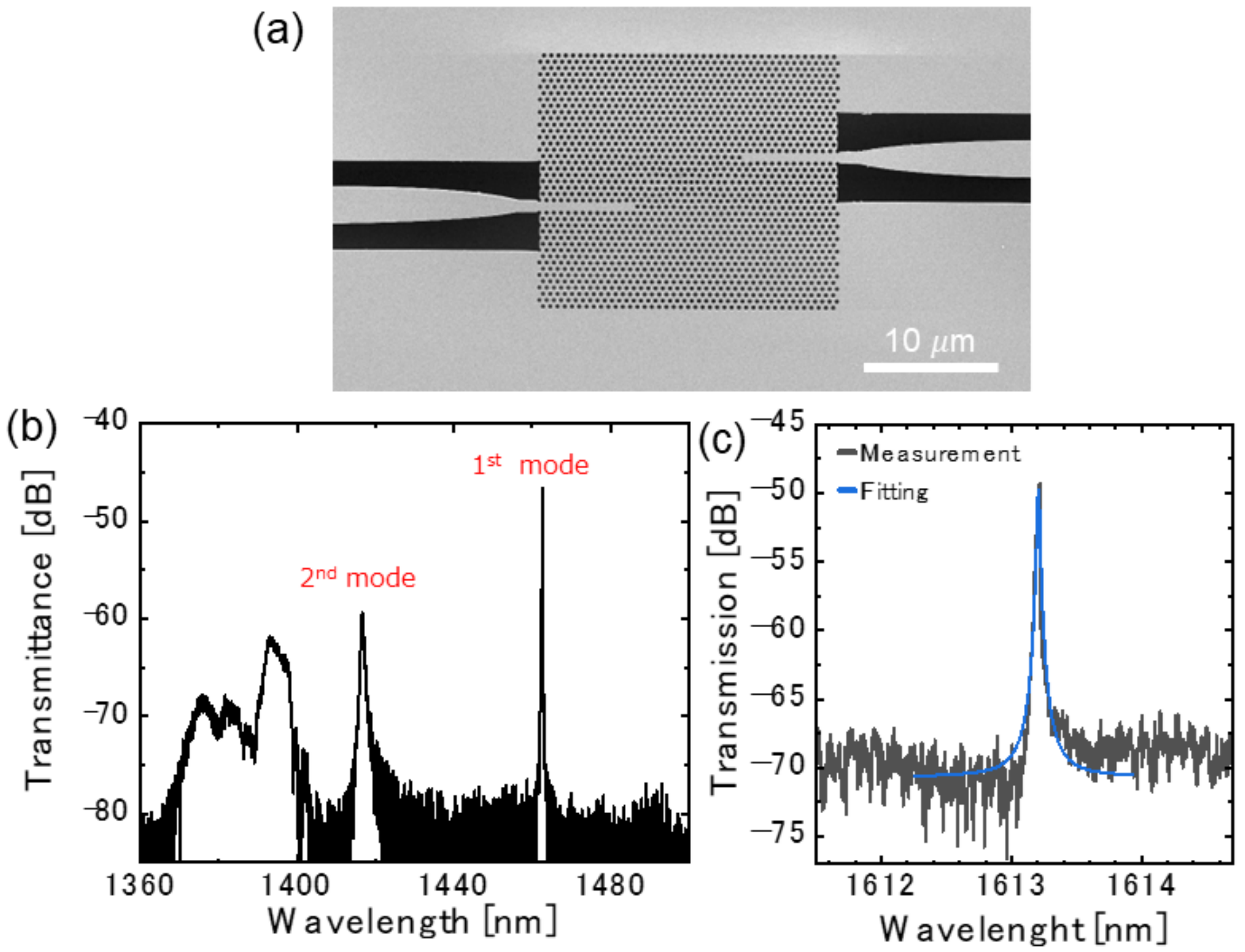

Fig. 3. **Fabricated photonic crystal cavity and its transmission spectra.**
(a) SEM image of the SiN photonic crystal cavity. (b) Broadband transmission spectrum. (c) Transmission spectrum of the cavity resonance for the d6-2 waveguide–cavity coupling configuration.

## Second and Third Harmonic Generation

As shown in Fig. 5 (a), third-harmonic generation (THG) was observed when the fundamental cavity mode was resonantly excited. For the measurements, a fs pulsed laser with an input power of 10 dBm was coupled into the device through the waveguide, and the generated SHG and THG signals were collected from the top surface using a 100× objective lens (Supplementary Note S3). SiN is a typical third-order nonlinear material, and THG has been widely demonstrated in ultrahigh-Q SiN microring resonators[46] and one-dimensional multilayered structures[47]. In contrast, THG from two-dimensional SiN PhC cavities has not, to our knowledge, been reported previously. The observation of THG indicates the strong optical field enhancement achieved in the inverse-designed cavity.

As indirect evidence, THG was not observed when higher-order cavity modes were excited (Supplementary Note S5). Since the higher-order modes exhibit *Q*/*V* values more than two orders of magnitude smaller than that of the fundamental mode (Under the d5 coupling condition, the simulated Q/V ratios of the fundamental and second cavity modes are $27{,}000(\lambda/n)^{-3}$ and $660(\lambda/n)^{-3}$, respectively), the intracavity field enhancement is insufficient to generate detectable THG. Furthermore, THG was reproducibly observed in several other

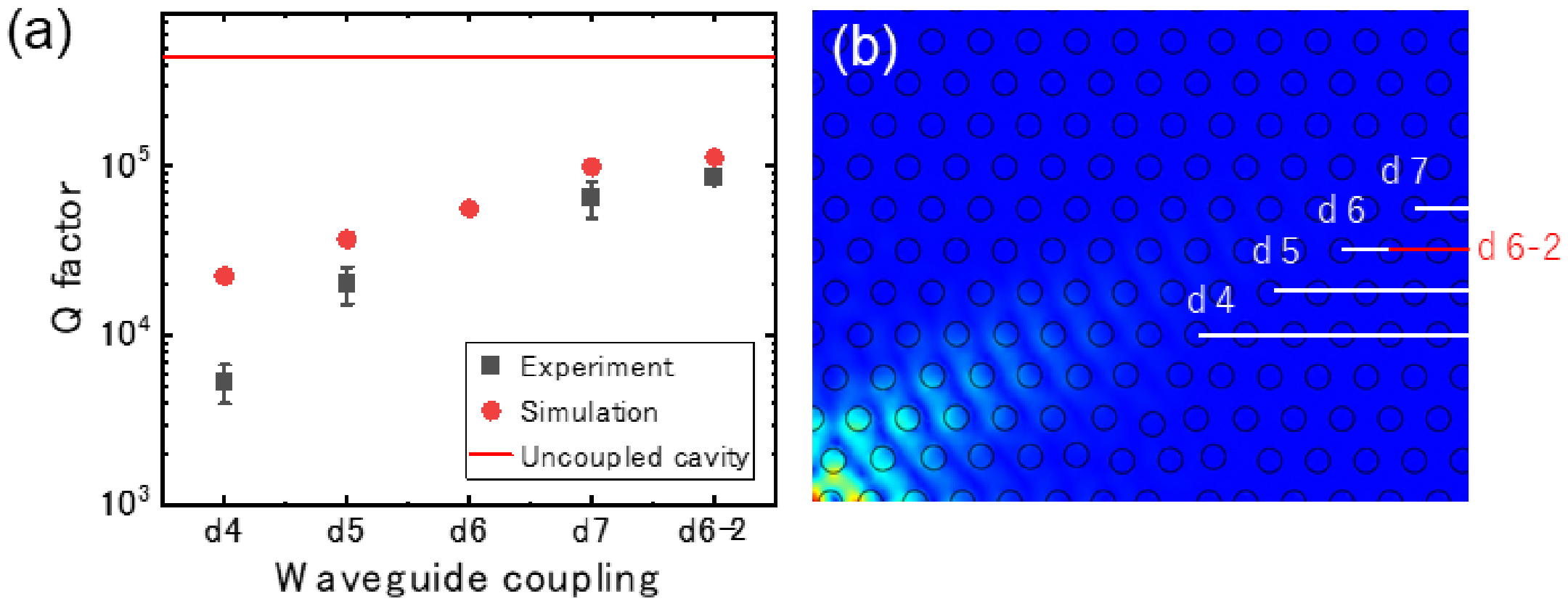

Fig. 4. **Dependence of the cavity quality factor on waveguide coupling.**
(a) Measured and simulated quality factor (Q) as a function of the waveguide coupling strength. For each coupling condition, 3–5 samples were evaluated, and the error bars represent the standard deviation. The red line indicates the simulated cavity Q without coupled waveguides. (b) Schematic showing the positions of the input and output waveguides.

high-*Q* cavities fabricated on the same chip (Supplementary Note S5), confirming that the nonlinear wavelength conversion originates from the optimized cavity design rather than from an individual device.

More interestingly, as shown in Fig. 5(a), SHG was also observed from the same cavity. Although amorphous SiN is generally regarded as a centrosymmetric material with negligible bulk $\chi^{(2)}$, symmetry breaking at the air–SiN interface is known to induce an effective surface $\chi^{(2)}$ [48,49,50]. Because the electric field of the L6/5 cavity is strongly localized around the PhC air holes, the optical mode efficiently drives the surface nonlinear polarization, enabling observable SHG. This field localization provides a large field-weighted interface-to-volume ratio, enhancing the relative contribution of the surface nonlinearity with respect to the bulk response. The effective nonlinear dipole moments responsible for SHG and THG, obtained by integrating the local nonlinear polarization density over the symmetry-broken surface layer and the SiN volume, respectively, can be approximately estimated as $p_{\mathrm{SHG}} \propto {\chi_{\mathrm{s}}}^{(2)} S\Gamma E_{\perp}^{2}$, and $p_{\mathrm{THG}} \propto \chi^{(3)} V E^{3}$, where $E_{\perp}$and $E$ are the electric-field component normal to the air-hole surface and the electric-field amplitude, respectively, obtained from FEM simulations. $V$ is the mode volume, $S$ is the total surface area of the air holes, $\Gamma$ is the overlap factor between the optical field and the hole surfaces, and ${\chi_{\mathrm{s}}}^{(2)}$ is the effective surface second-order susceptibility, defined as ${\chi_{\mathrm{s}}}^{(2)} = \chi^{(2)} \Delta s$ where $\Delta s$ denotes the effective thickness of the symmetry-broken surface layer, estimated to be approximately 0.3 nm, corresponding approximately to a single atomic (or molecular) layer, following the conventional surface SHG model[11,51]. Using literature values[52,53] of $\chi^{(2)} \sim 2.5 \times 10^{-12}$ m/V and $\chi^{(3)} \sim 6 \times 10^{-20}$ $m^2/V^2$, together with $\Gamma$ of a few tens percent estimated from the simulated field distribution and $E/E_{\perp} \approx 10$ (Supplementary Note S5), the SHG and THG intensities can be estimated as the squares of the corresponding effective nonlinear dipole moments as a function of the input power, as shown in Fig. 5(b). We therefore superimposed the experimentally measured THG-to-SHG peak intensity ratios at different pump powers onto the calculated dependence. Although the estimated ratio differs from the experimental value by a factor of a few, the calculation does not account for wavelength-dependent collection efficiencies of the objective lens or polarization-dependent detection efficiencies for SHG and THG. Considering these experimental factors, the agreement is reasonably good, supporting the interpretation that the observed SHG originates from surface-induced second-order nonlinearity, whereas THG is generated through the intrinsic bulk third-order nonlinearity of SiN.

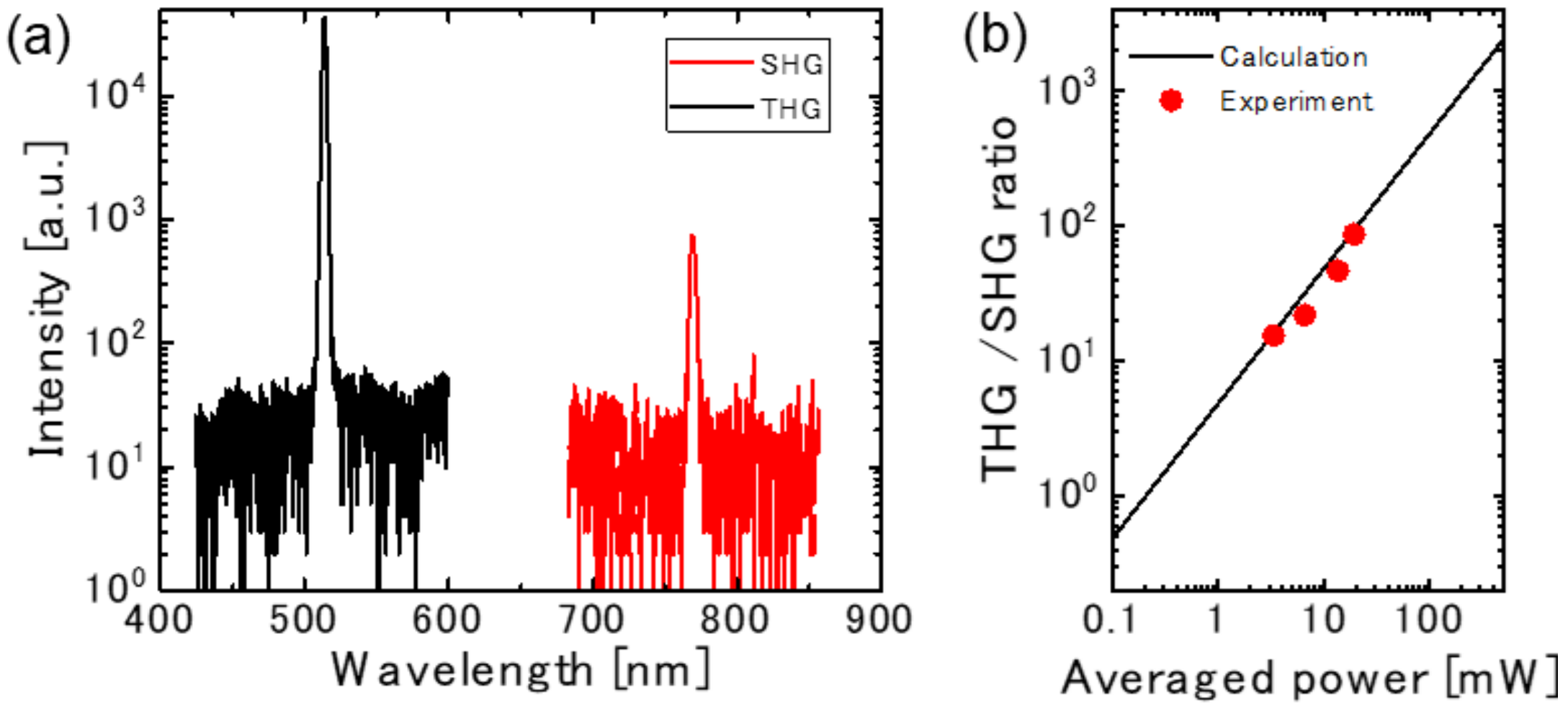

Fig. 5. **Characterization of second- and third-harmonic generation.** (a) Second- and third-harmonic generation (SHG and THG) spectra obtained from the same photonic crystal cavity. (b) Excitation power dependence of the THG/SHG ratio. The input power was measured before coupling into the waveguide.

## Discussion

In this work, we demonstrated that inverse design enables the realization of a high-*Q*/*V* two-dimensional PhC cavity on a low-index SiN platform. By optimizing an L6/5 cavity using a parallel derivative-free optimization algorithm, we achieved a simulated quality factor exceeding $4\times10^5$ with a mode volume of approximately $0.96(\lambda/n)^3$. The fabricated devices experimentally exhibited *Q*-factors approaching $8\times10^4$, which, to the best of our knowledge, represents the highest reported value for a two-dimensional SiN PhC cavity fabricated from stoichiometric SiN. To quantitatively assess the performance of our cavity in the context of previous studies, Table 1 compares the simulated and experimentally measured *Q* factors, mode volumes, and *Q*/*V* values of representative SiN PhC cavities. Our cavity exhibits one of the highest Q/V values among Lx-type PhC cavities based on stoichiometric SiN. Notably, its *Q*/*V* is also comparable to those reported for high-*Q* cavities based on Si-rich SiN, despite the lower refractive index of stoichiometric SiN. The higher refractive index of Si-rich SiN provides stronger optical confinement, whereas achieving simultaneously high *Q* and small *V* is more challenging in low-index stoichiometric SiN. These results therefore highlight the effectiveness of inverse design in overcoming, at least in part, the limitations imposed by the relatively low refractive-index contrast of stoichiometric SiN.

| Ref | Cavity type | n | Depo. | Vertical structure | λ | Sim. Q | Exp. Q | V $[(\lambda/n)^3]$ | Q/V $[(\lambda/n)^{-3}]$ |
|---|---|---|---|---|---|---|---|---|---|
| [28] | 2D, L3 | 2.01 | LPCVD | Air-bridge | 550–800 nm | $4.7 \times 10^3$ | $1.46 \times 10^3$ | 1.32 | $1.11 \times 10^3$ |
| [27] | | 2.1 | N/R | Air-bridge | 650 nm | — | $1.5 \times 10^3$ | N/R | — |
| [26] | | 2 | — | $SiO_2$ bottom cladding | 717 nm | $2.5 \times 10^3$ | — | N/R | — |
| [54] | | 2 | LPCVD | Air-bridge | 520 nm | $3.8 \times 10^4$ | $3.0 \times 10^3$ | N/R | — |
| This work | 2D, L6/5 | 2.0 | PECVD | Air-bridge | Telecom | $> 4 \times 10^5$ | $8.0 \times 10^4$ | 0.96 | $8.33 \times 10^4$ |
| [29] | 2D, line-width-modulated | Silicon-rich 2.54 | PECVD | Air-bridge | Telecom | $5.2 \times 10^5$ | $1.2 \times 10^5$ | 0.77 | $1.56 \times 10^5$ |
| [30] | | | PECVD | Air-bridge | Telecom | $2.1 \times 10^4$ | $1.3 \times 10^4$ | 0.73 | $1.78 \times 10^4$ |
| [32] | 1D nanobeam | 2 | LPCVD | $SiO_2$/PMMA | 740 nm | $1 \times 10^5$ | $7.0 \times 10^3$ | 2.5 | $2.8 \times 10^3$ |
| [31] | | 2.0 | LPCVD | $SiO_2$-clad | Telecom | $2.8 \times 10^7$ | $4.4 \times 10^5$ | N/R | — |

Table. 1 **Comparison of SiN photonic crystal cavities**
Comparison of representative SiN photonic crystal cavities reported in previous studies. Experimental Q/V values were calculated using the maximum experimentally measured Q factors and the simulated mode volumes reported in the corresponding studies. N/R denotes not reported.

The significance of this work extends beyond achieving a high *Q*-factor. Compared with silicon, SiN offers a broad transparency window spanning the visible to the telecommunications wavelength range, negligible two-photon absorption at telecommunication wavelengths, and full CMOS compatibility. These properties make SiN an attractive platform for nonlinear and quantum photonics. However, its relatively low refractive index has limited optical confinement, making it difficult to realize PhC cavities with sufficiently large $Q/V$. Our results demonstrate that inverse design can effectively overcome this limitation, providing a practical route toward high-performance SiN nanocavities. Very recently, inverse-designed SiN photonic crystal cavities[54] optimized for simultaneous *Q*-factor and improved free-space coupling have also been demonstrated using alternative optimization strategies, further highlighting the growing importance of inverse-design approaches for low-index photonic crystal platforms.

In the present study, the inverse-design process was restricted to optimizing the positions of circular air holes. Further improvements may be possible by extending the design degrees of freedom to include the sizes and shapes of the air holes or, more generally, by employing topology optimization without restricting the geometry to circular holes. Previous topology-optimization studies have demonstrated that such increased geometrical freedom can lead to unconventional cavity structures, including dielectric bowtie cavities surrounded by elliptic ring-like gratings, enabling strong optical confinement and enhanced $Q/V$ [55] . More recently, fabrication-constrained topology optimization has enabled the experimental realization of such dielectric bowtie cavities[56] . These results suggest that incorporating non-circular or arbitrary-shaped features into the inverse-design framework may provide an additional route toward further improving the performance of SiN PhC cavities.

The observation of both third-harmonic generation and surface-induced second-harmonic generation further confirms the strong field confinement achieved in the optimized cavity. While THG originates from the intrinsic third-order nonlinearity of SiN, SHG is attributed to the symmetry breaking at the air–SiN interfaces, where the optical field is strongly localized around the PhC holes. This field localization enhances the relative contribution of the surface nonlinear response compared with the bulk nonlinear response, allowing both contributions to become experimentally accessible within the same nanocavity. The experimental observation of both nonlinear processes, together with the agreement between the measured THG/SHG intensity ratio and a simple theoretical estimation, provides strong evidence that the inverse-designed cavity achieves exceptionally large field enhancement. In the present design, only the fundamental mode is resonantly enhanced, whereas essentially no resonant enhancement occurs at the SHG or THG wavelengths. As discussed in previous studies of nonlinear photonic crystal cavities[57] [58] , substantially higher conversion efficiencies could, in principle, be achieved by employing a doubly resonant cavity that simultaneously confines the fundamental and harmonic modes. For a relatively low-index material such as SiN, simultaneously achieving resonances at widely separated wavelengths may require careful design; however, inverse design provides a promising approach for realizing such structures. In future designs, the fundamental and SHG/THG wavelengths could be simultaneously incorporated into the objective function, potentially enabling multi-resonant SiN PhC cavities optimized for nonlinear frequency conversion.

The demonstrated platform opens new opportunities for integrated nonlinear photonics in the visible and near-infrared wavelength ranges. Because SiN is compatible with a wide variety of functional materials, including GaN, ZnO, diamond, rare-earth emitters, and two-dimensional materials, the present cavity architecture provides a promising platform for efficient wavelength conversion, enhanced light–matter interaction, integrated quantum photonics, and future heterogeneous photonic integrated circuits.

## Method

### Cavity optimization

Recently, inverse-design techniques have emerged as powerful tools for improving the performance of a wide variety of photonic devices. In particular, inverse design has enabled the realization of PhC cavities with exceptionally large $Q/V$ ratios. Among the most successful approaches are machine-learning-based optimization and gradient-descent methods using automatic differentiation. Recent studies have also experimentally demonstrated the effectiveness of AI-assisted approaches for photonic device design and optimization[59] . However, the former generally requires the preparation of training datasets, whereas the latter relies on dedicated simulation frameworks that support automatic differentiation. In contrast, derivative-free optimization methods can be readily implemented in virtually any simulation environment, making them attractive for the rapid prototyping of PhC devices.

The parallel simplex algorithm (PSA) is one of the simplest and most compact derivative-free optimization methods. Its major drawback, however, is the large computational cost associated with high-dimensional optimization. To overcome this limitation, we previously implemented a computationally parallelized optimization framework leveraging a parallel simplex algorithm, thereby significantly reducing the computation time required for the objective convergence of inverse-designed PhC cavities. Using this approach, we successfully optimized the cavity figure of merit in nanowire cavities in our previous work[40]. In the present work, we apply the same optimization framework to maximize the $Q/V$ ratio of a two-dimensional SiN photonic crystal cavity. In addition, we also optimized the cavity using the particle swarm optimization (PSO) algorithm as an alternative inverse-design approach[60]. The details are provided in the Supplementary Note S1. Interestingly, although different optimization algorithms converge to distinct hole-position modulation patterns, they produce comparable $Q/V$ ratios. This indicates that the parameter space contains numerous local optima with similar performance. However, in practical fabrication, connections to input and output waveguides are required. Since the hole modulation optimized by PSO extends over a relatively broad region and may slightly affect the input and output waveguides, we adopted the PSA design.

### Device fabrication

The overall fabrication process is schematically illustrated in Supplementary Note S2. A 425-nm-thick SiN film was deposited on a silicon substrate by plasma-enhanced chemical vapor deposition (PECVD). A resist layer was then spin-coated, and the photonic crystal cavity and waveguide patterns were defined by electron-beam lithography followed by resist development. The SiN layer was subsequently patterned by inductively coupled plasma (ICP) dry etching until the underlying silicon substrate was exposed. After removing the resist, the silicon substrate was selectively undercut by potassium hydroxide (KOH) wet etching to form an air-bridge structure. Finally, the wafer was diced into individual chips. Because the lateral etch rate of silicon in KOH is relatively slow, prolonged wet etching is required to fully release the air-bridge structure. During this process, the SiN layer is also slightly thinned. Therefore, the initial SiN thickness was intentionally increased to 425 nm so that the final device thickness after wet etching was approximately 350 nm, which was the target thickness for the cavity design.

## Data availability

The datasets generated and/or analyzed during the current study are not publicly available due to the large volume and instrument-specific formats of the raw experimental and simulation data, but are available from the corresponding author on reasonable request.

## Acknowledgments

The authors gratefully acknowledge the contributions of Shinichi Fujiura to the device measurements and Hidenori Onji to the preparation of the CAD layouts. They further thank Yoshio Ohki, Toshifumi Watanabe, Mizuki Ikeya and Osamu Moriwaki of NTT Advanced Technology Corporation, as well as Junichi Asaoka of NTT Devices Cross Technologies Corporation, for their technical expertise in SEM characterization, electron-beam lithography, and the fabrication of the photonic devices.

## Funding

This work has been supported by the JSPS KAKENHI Grant Numbers 23K26581 and 26K01425.

## Author contributions

M.T. conceived the idea and designed the study. P.H. and X.L. optimized the PhC cavities. M.T., S.Y., and J.Z. performed the measurements. S.Y. estimated the waveguide coupling. M.T. and K.N. analyzed the nonlinear effects. M.T. fabricated the devices with the help of M.O. and T.A. The project was coordinated by M.T., H.S., and M.N. All authors contributed to the manuscript and scientific discussions.

## Competing interests

The authors declare no competing interests.

## Additional information

The online version contains supplementary material available at XXXXXXX.

# Supplemental: Inverse-Designed High-Q and Small-Mode-Volume Silicon Nitride Photonic Crystal Cavities for Second- and Third-Harmonic Generation

**M. Takiguchi[1,2], P. Heidt[2], X. Z. Lim[2], J. Zöllner[2], S. Yanagimoto[2], K. Nakayama[3], T. Aihara[4], M. Ono[1,2], H. Sumikura[1,2], and M. Notomi[1,2,5]**

*[1]NTT Nanophotonics Center, NTT, Inc., 3-1 Morinosato Wakamiya, Atsugi, Kanagawa 243–0198, Japan*

*[2]Basic Research Laboratories, NTT, Inc., 3-1 Morinosato Wakamiya, Atsugi, Kanagawa 243–0198, Japan*

*[3]International Center for Elementary Particle Physics, University of Tokyo, Bunkyo-ku, Tokyo 113-0033, Japan.*

*[4]Device Technology Laboratories, NTT, Inc., 3-1 Morinosato Wakamiya, Atsugi, Kanagawa 243–0198, Japan*

*[5]Department of Physics, Institute of Science Tokyo, Meguro-ku, Tokyo 152-8551, Japan*

*Author e-mail address:* masato.takiguchi@ntt.com

## S1: Inverse design

Because of the relatively low refractive index of SiN, the photonic bandgap is relatively narrow, making it difficult to realize ultrahigh-*Q* photonic crystal (PhC) cavities using conventional deterministic optimization methods. To overcome this limitation, we employed inverse design to maximize the *Q*/*V* ratio of a SiN-based two-dimensional PhC cavity. In this work, all inverse-design calculations were performed on a workstation equipped with two Intel Xeon Gold 6258R processors (2.70 GHz, 28 cores per processor; 56 physical cores in total) and 1.5 TB of DDR4 memory.

Figure S1 compares the optimized hole-position modulations obtained using the parallel simplex algorithm (PSA) [Ref.1] and particle swarm optimization (PSO) [Ref.2]. As shown in Fig. S1, the two optimization methods produce substantially different hole-position modulations. Nevertheless, both optimized structures

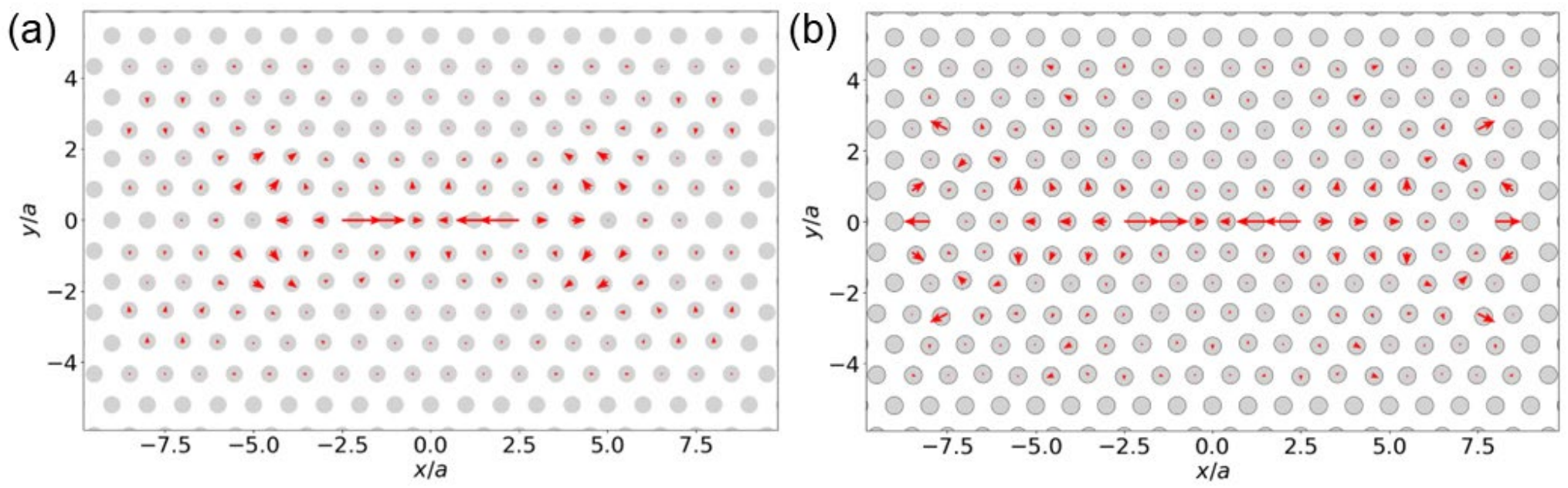

Fig. S1. **Hole-position shifts obtained using different optimization methods.**
(a) Hole-position modulation of the photonic crystal optimized using the parallel simplex algorithm. (b) Hole-position modulation of the photonic crystal optimized using the particle swarm optimization (PSO) algorithm. The coordinates are normalized to the lattice constant, *a*.

parallel simplex algorithm

| Hole No. | x | y | Hole No. | x | y |
|---|---|---|---|---|---|
| 1 | 0.404423 | 0 | 28 | 0.490529 | 2.61277 |
| 2 | 1.230634 | 0 | 29 | 1.488424 | 2.58671 |
| 3 | 2.117736 | 0 | 30 | 2.49533 | 2.576533 |
| 4 | 3.105409 | 0 | 31 | 3.494642 | 2.58953 |
| 5 | 4.127722 | 0 | 32 | 4.45675 | 2.618462 |
| 6 | 5.020677 | 0 | 33 | 5.443759 | 2.594514 |
| 7 | 6.054331 | 0 | 34 | 6.461004 | 2.543803 |
| 8 | 7.030797 | 0 | 35 | 7.488578 | 2.534765 |
| 9 | 7.999685 | 0 | 36 | 8.511448 | 2.534783 |
| 10 | 0.506988 | 0.947794 | 37 | 0 | 3.47929 |
| 11 | 1.519859 | 0.912367 | 38 | 1.005714 | 3.463149 |
| 12 | 2.541294 | 0.875937 | 39 | 2.010039 | 3.433786 |
| 13 | 3.520191 | 0.929617 | 40 | 3.032595 | 3.435198 |
| 14 | 4.424491 | 0.963786 | 41 | 4.030561 | 3.459291 |
| 15 | 5.435351 | 0.942507 | 42 | 5.01949 | 3.463523 |
| 16 | 6.488999 | 0.915926 | 43 | 6.013541 | 3.424888 |
| 17 | 7.506337 | 0.90907 | 44 | 7.00193 | 3.400355 |
| 18 | 8.493576 | 0.909115 | 45 | 7.999919 | 3.403255 |
| 19 | 0 | 1.7216 | 46 | 0.506845 | 4.32994 |
| 20 | 0.954936 | 1.71194 | 47 | 1.50548 | 4.329804 |
| 21 | 1.950436 | 1.688799 | 48 | 2.518475 | 4.330373 |
| 22 | 2.954871 | 1.713221 | 49 | 3.540293 | 4.329735 |
| 23 | 3.922429 | 1.789142 | 50 | 4.541832 | 4.330555 |
| 24 | 4.894189 | 1.798039 | 51 | 5.541647 | 4.330055 |
| 25 | 5.938396 | 1.761357 | 52 | 6.523957 | 4.329893 |
| 26 | 6.976832 | 1.746761 | 53 | 7.503786 | 4.329974 |
| 27 | 8.00001 | 1.759454 | 54 | 8.496218 | 4.330486 |

particle swarm optimization

| Hole No. | x | y | Hole No. | x | y |
|---|---|---|---|---|---|
| 1 | 0.393309 | 0 | 28 | 0.489919 | 2.564181 |
| 2 | 1.218777 | 0 | 29 | 1.484948 | 2.580217 |
| 3 | 2.145492 | 0 | 30 | 2.516802 | 2.631096 |
| 4 | 3.130549 | 0 | 31 | 3.535826 | 2.634011 |
| 5 | 4.116189 | 0 | 32 | 4.520155 | 2.644653 |
| 6 | 5.108827 | 0 | 33 | 5.548947 | 2.598003 |
| 7 | 6.047963 | 0 | 34 | 6.516364 | 2.660106 |
| 8 | 6.970333 | 0 | 35 | 7.66776 | 2.682265 |
| 9 | 8.248706 | 0 | 36 | 8.497704 | 2.620162 |
| 10 | 0.463897 | 0.866025 | 37 | 0 | 3.514979 |
| 11 | 1.480283 | 0.914498 | 38 | 1.005409 | 3.426489 |
| 12 | 2.533129 | 0.934572 | 39 | 1.990816 | 3.449814 |
| 13 | 3.514833 | 0.963321 | 40 | 3.023252 | 3.499526 |
| 14 | 4.537503 | 0.953466 | 41 | 4.07158 | 3.503754 |
| 15 | 5.49559 | 0.989168 | 42 | 5.021795 | 3.475326 |
| 16 | 6.460821 | 0.858853 | 43 | 5.99621 | 3.494424 |
| 17 | 7.45827 | 0.887616 | 44 | 6.997595 | 3.470515 |
| 18 | 8.381369 | 0.950609 | 45 | 8.001604 | 3.508419 |
| 19 | 0 | 1.745025 | 46 | 0.511539 | 4.334572 |
| 20 | 0.974211 | 1.739194 | 47 | 1.461331 | 4.346818 |
| 21 | 1.995189 | 1.746629 | 48 | 2.499599 | 4.379765 |
| 22 | 3.00328 | 1.768496 | 49 | 3.531744 | 4.304687 |
| 23 | 4.005977 | 1.753189 | 50 | 4.569429 | 4.358044 |
| 24 | 5.005467 | 1.746483 | 51 | 5.505066 | 4.355565 |
| 25 | 6.073912 | 1.763102 | 52 | 6.49231 | 4.304395 |
| 26 | 7.073985 | 1.660617 | 53 | 7.474889 | 4.334427 |
| 27 | 7.981486 | 1.748378 | 54 | 8.460238 | 4.342445 |

Table S1. **Details of the photonic crystal hole-position shifts.**
Positions of the photonic crystal holes in the first quadrant used in the simulations. The hole positions are normalized by the lattice constant, and the hole diameter is fixed at half the lattice constant. The left and right columns show the results obtained using the parallel simplex algorithm and particle swarm optimization, respectively.

share a common feature: the largest hole displacements are concentrated around the cavity center, indicating that this region plays the dominant role in suppressing radiation loss and enhancing the cavity *Q*. The distinct optimized geometries obtained using PSA and PSO suggest that the optimization landscape contains a large number of local optima. Despite the different hole-position distributions, however, the optimized cavity performances are remarkably similar. The PSA-optimized cavity exhibits a quality factor and mode volume of $4.4 \times 10^5$ and $0.96(\lambda/n)^3$, respectively, whereas the corresponding values obtained using PSO are $1.7 \times 10^5$ and $0.94(\lambda/n)^3$, demonstrating comparable resulting *Q*/*V* ratios. The existence of multiple local optima is further supported by the dependence of the optimization on the initial cavity geometry. In this study, the optimization was primarily performed using an L6/5 cavity as the initial structure. Although the L6/5 and L4/3 cavities have different initial geometries, they contain the same number of air holes and therefore share the same parameter space for hole-position optimization. Therefore, theoretically, both initial structures should also possess the same global optimum. However, optimization starting from the L4/3 cavity converged to a different solution, yielding a quality factor and mode volume of $1.9 \times 10^5$ and $1.0(\lambda/n)^3$, respectively. Together with the comparison between PSA and PSO, these results indicate that the high-dimensional parameter space of SiN photonic crystal cavities contains numerous local optima with comparable $Q/V$ values. The optimized hole-position parameters obtained using both methods are summarized in Table S1.

We also investigated the robustness of the optimized cavity against deviations from the designed structural parameters, taking into account variations that may occur during actual fabrication. Because the hole positions are defined by electron-beam lithography, a large deviation in the position of only a single specific hole is unlikely to occur. In contrast, systematic deviations in the slab thickness and overall air-hole diameter from their designed values can readily arise from variations in the wet- and dry-etching conditions, respectively. We therefore evaluated the cavity Q factor as a function of the air-hole diameter and slab thickness, as shown in Fig. S2. The results show that the cavity maintains a high Q factor even when the air-hole diameter deviates from the designed value by approximately ±15 nm or the slab thickness deviates by approximately ±30 nm, demonstrating that the optimized cavity is robust against realistic variations in both the air-hole diameter and slab thickness.

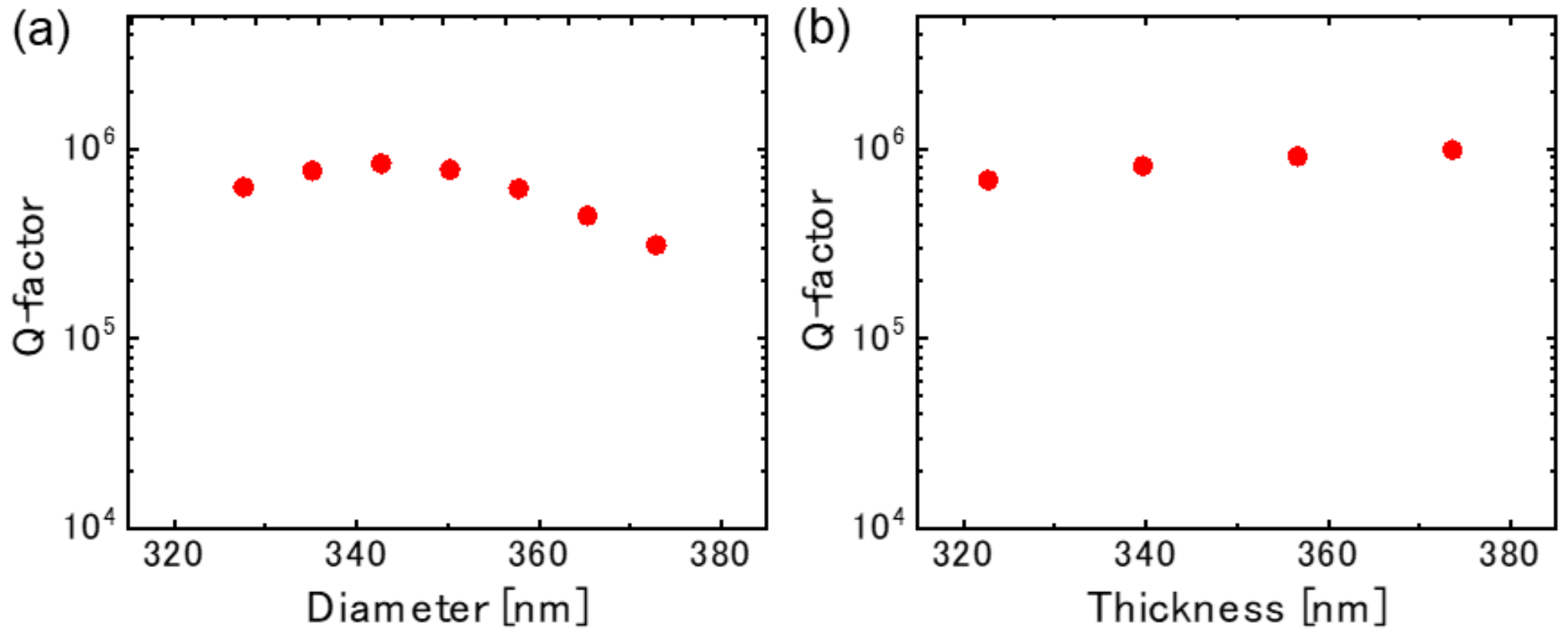

Fig. S2. **Dependence of the Q factor on hole diameter and slab thickness.**
(a) Hole-diameter dependence of the Q factor. (b) Slab-thickness dependence of the Q factor

## S2: Fabrication

The fabrication process is schematically illustrated in Supplementary Fig. S3. A 425-nm-thick SiN film was deposited on a silicon substrate by plasma-enhanced chemical vapor deposition (PECVD). To match the material parameters used in the inverse-design simulations, the deposition conditions were optimized to obtain a refractive index of approximately 2.0. The measured refractive index obtained by spectroscopic ellipsometry is shown in Supplementary Fig. S3 (b). PECVD was also selected because it produces SiN films with relatively low residual stress compared with other deposition techniques, such as low-pressure chemical vapor deposition (LPCVD), electron cyclotron resonance chemical vapor deposition (ECR-CVD), and sputtering. A resist layer was subsequently spin-coated, and the photonic crystal cavity and waveguide patterns were defined by electron-beam lithography followed by resist development. The SiN layer was then patterned by inductively coupled plasma (ICP) dry etching until the underlying silicon substrate was exposed. After removing the resist, the silicon substrate was selectively undercut by potassium hydroxide (KOH) wet etching to form an air-bridge structure. During the wet-etching process, we found that very large suspended structures, such as W8 photonic crystal waveguides, exhibited slight bending caused by residual stress in the SiN film. However, this deformation was sufficiently small that no noticeable degradation of the optical characteristics was observed. Finally, the wafer was diced into individual chips.

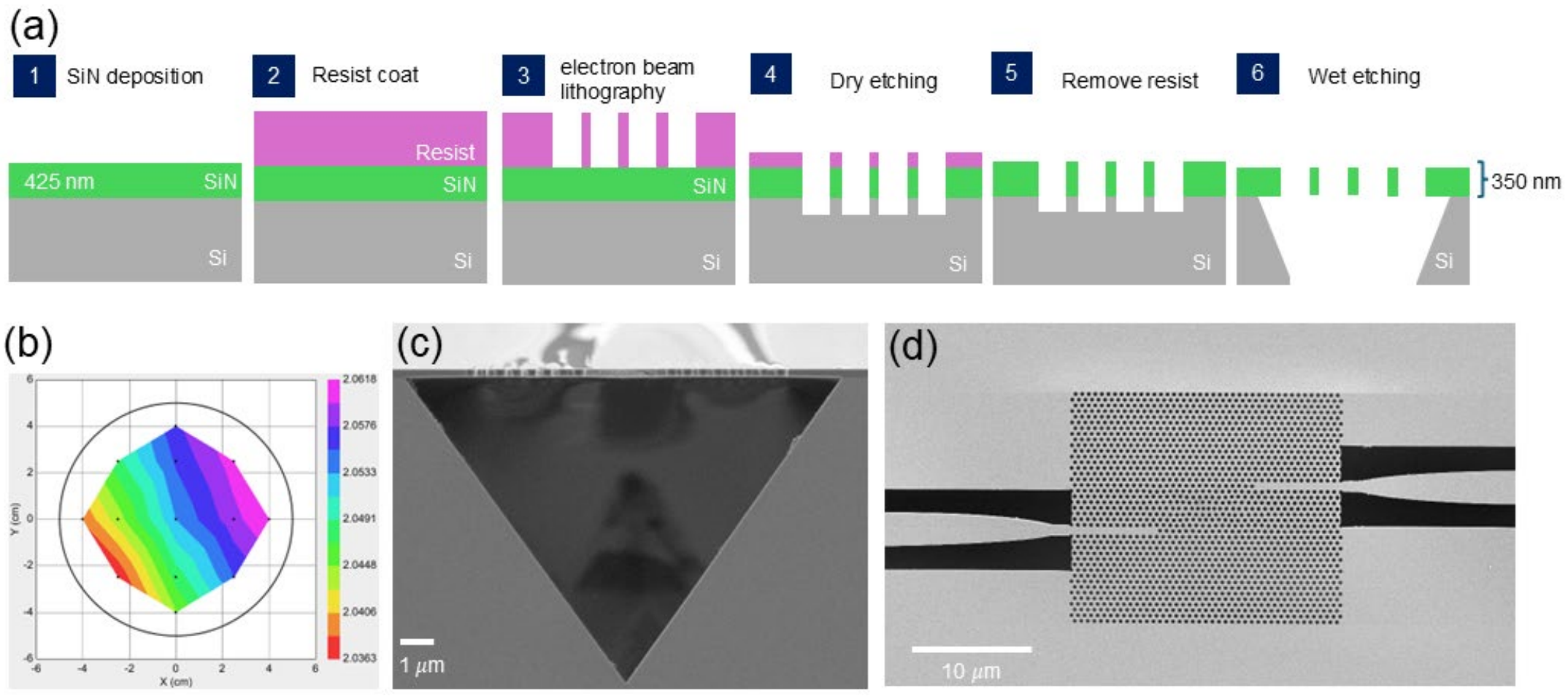

Fig. S3. **Fabrication process of the SiN photonic crystal cavity.**
(a) Schematic illustration of the fabrication process flow. (b) Refractive index dispersion measured by spectroscopic ellipsometry. (c) Cross-sectional scanning electron microscope (SEM) image of the photonic crystal waveguide after wet etching. (d) SEM image of the fabricated photonic crystal cavity.

Because the lateral etch rate of silicon in KOH is relatively slow, prolonged wet etching is required to completely release the air-bridge structure. During this process, the SiN layer is also slightly thinned. Therefore, the initial SiN thickness was intentionally increased to 425 nm so that the final device thickness after wet etching was approximately 350 nm, which corresponds to the target thickness used in the cavity design.

## S3: Measurement setup

Figure S4 shows a schematic of the experimental setup. The cavity transmission spectrum was obtained by sweeping the wavelength of a tunable continuous-wave (CW) laser. The input light was coupled into the chip through the edge facet and guided to the cavity via an on-chip waveguide. For the second- and third-harmonic generation (SHG and THG) measurements, an 80 MHz repetition-rate femtosecond pulsed laser was used as the excitation source. The generated SHG and THG signals were collected through an objective lens positioned above the sample and analyzed using a spectrometer (300 grooves/mm blaze grating) equipped with a liquid-nitrogen-cooled CCD detector.

As shown in Fig. 5 (a), the measured SHG and THG spectra are considerably broader than the cavity resonance. This is mainly attributed to the broad spectral bandwidth (>10 nm) of the femtosecond excitation pulses and the finite spectral resolution of the spectrometer (approximately 0.2–0.4 nm). The broad bandwidth of the excitation pulse is transferred to the generated harmonic signals through the nonlinear frequency-conversion process. In addition, nonlinear effects may induce temporal shifts of the cavity resonance, which could further contribute to the observed spectral broadening.

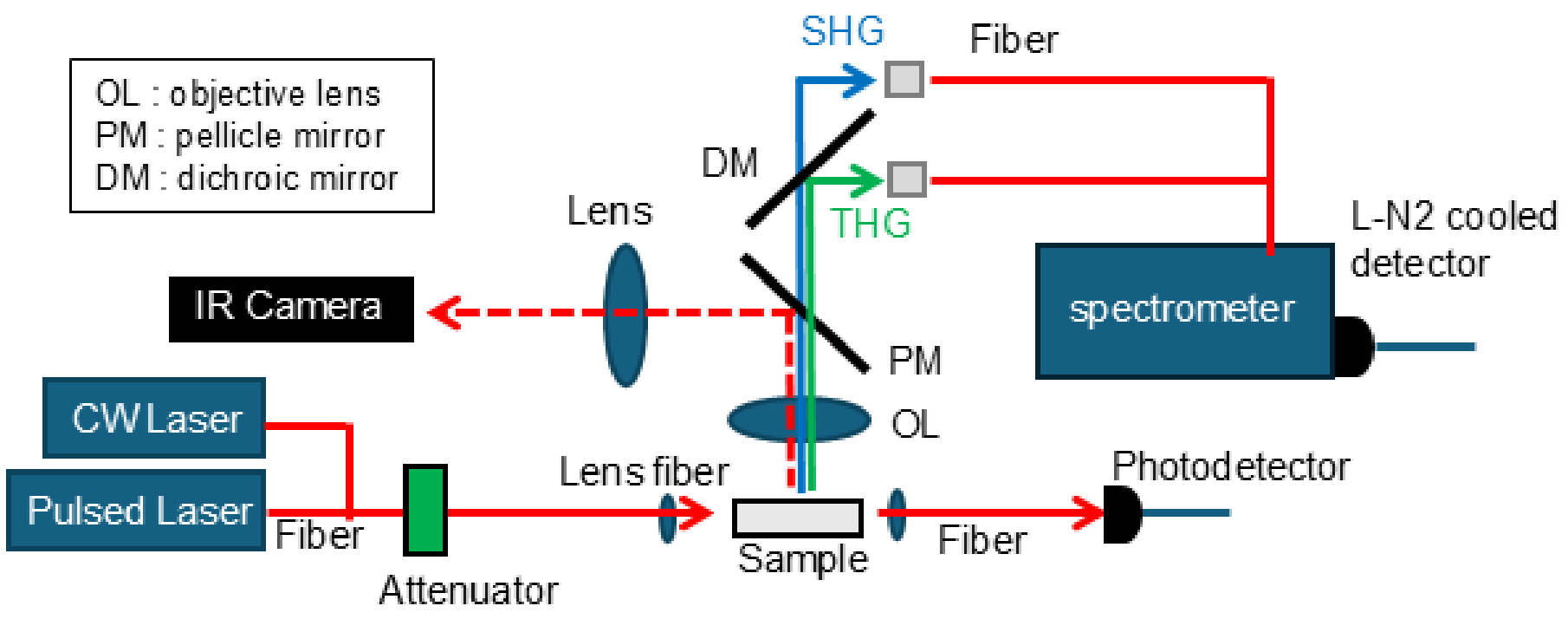

Fig. S4. **Measurement setup.**
Schematic illustration of the experimental setup used for transmission measurements and second-harmonic generation (SHG) and third-harmonic generation (THG) measurements.

## *S4*: SiN photonic crystal property

To characterize the fundamental properties of the SiN PhC, the transmission characteristics of a W1 PhC waveguide were first investigated. Figure S5(a) shows an optical microscope image of the fabricated device. As shown in the figure, the suspended region formed by KOH undercut etching appears with a different color contrast, indicating that the SiN slab is successfully released from the silicon substrate. A W1 PhC waveguide is located at the center of the device, while mode converters are integrated at both ends. The W1 waveguide is connected to the chip facets through W8 PhC waveguides. The insertion loss of the coupling system was evaluated using reference structures. The measured transmission through a reference W8 PhC waveguide exhibited a total insertion loss of approximately 15 dB, including the fiber-to-chip coupling losses at the input and output facets. For a structure consisting of W8 PhC waveguides, W8–W1 mode converters, and a W1 PhC waveguide, the total insertion loss was approximately 25 dB. Therefore, neglecting the propagation loss of the waveguides, the additional loss associated with the two W8–W1 mode converters is estimated to be approximately 10 dB in total, corresponding to approximately 5 dB per mode converter. These values correspond to samples with relatively good coupling conditions, and the actual insertion loss varies depending on factors such as the quality of the cleaved facets and the lattice constant of the PhC structure.

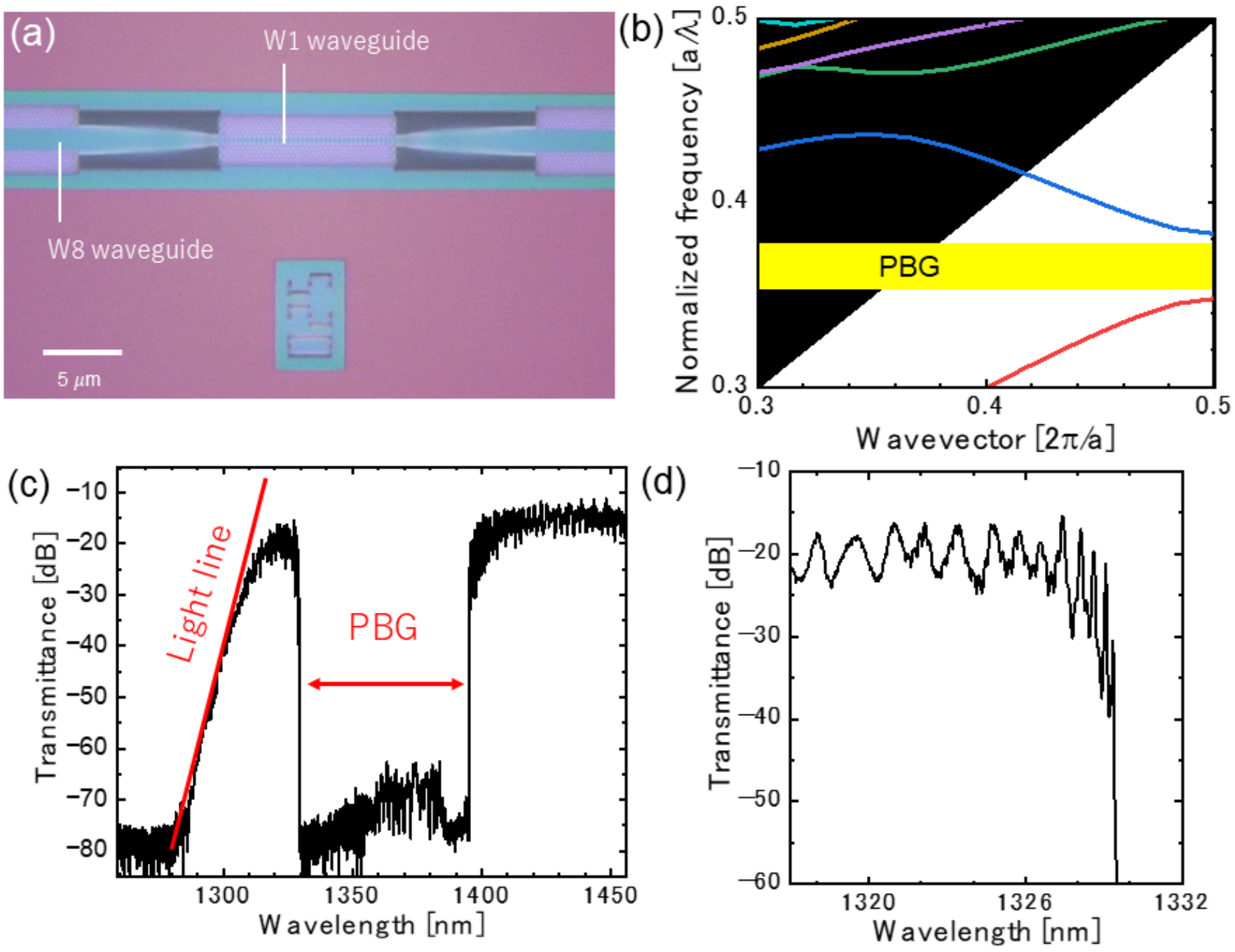

Fig. S5. **SiN photonic crystal waveguide.**
(a) Optical microscope image of the photonic crystal waveguide. (b) Photonic crystal band structure along the Γ–K direction. (c) Transmission spectrum. (d) Enlarged view of the transmission spectrum.

Figure S5(b) shows the calculated photonic band diagram of the SiN PhC waveguide along the Γ–K direction. A clear photonic bandgap is observed, confirming the existence of the guided mode within the designed wavelength range. Figures S5(c) and S5(d) show the measured transmission spectra of the W1 PhC waveguide. On the short-wavelength side, the transmission decreases because the guided mode lies above the light line, resulting in increased radiation loss. Near the photonic band edge, the interference fringes become more closely spaced in wavelength, indicating an increase in the group index associated with the slow-light effect [Ref.3].

In the main text, we presented the transmission spectrum for the d6-2 waveguide-coupling configuration. Here, we additionally present the transmission spectra for all the waveguide-coupling configurations in Fig S6.

[Ref. 3] M. Notomi, et.al., Phys. Rev. Lett., 87.253902 (2001)

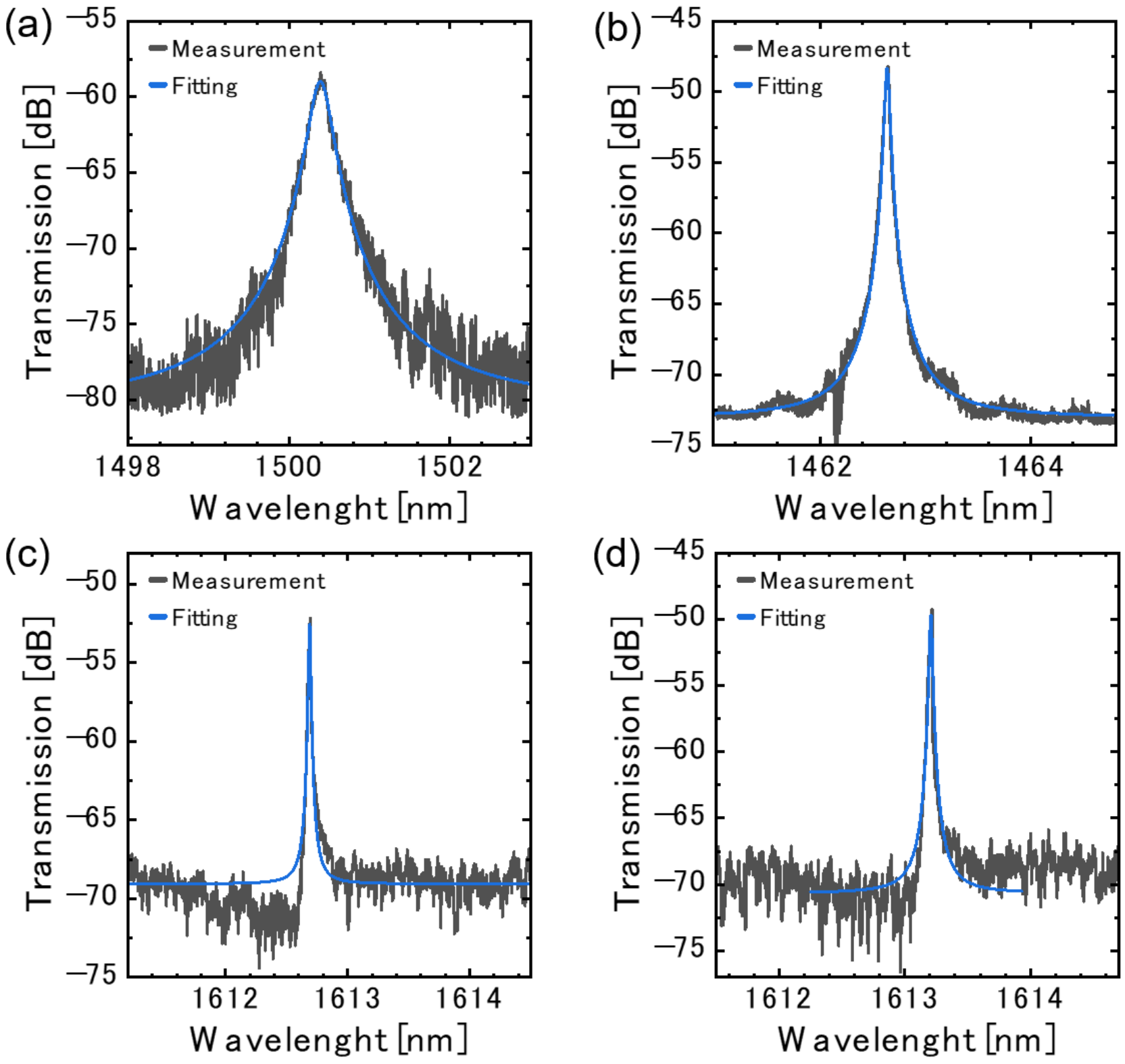

Fig. S6. **Measured cavity transmission spectra for different waveguide-coupling configurations.**
(a) d = 4, a = 680 nm, Q ≈ 5,270; (b) d = 5, a = 660 nm, Q ≈ 29,000; (c) d = 7, a = 740 nm, Q ≈ 81,900; and (d) d = 6-2, a = 740 nm, Q ≈ 87,800.

## S5: Third-harmonic generation

Our SiN PhC cavity exhibits efficient third-harmonic generation (THG) only when the fundamental cavity mode is excited. Figure S7 shows the cavity transmission spectrum (reproduced from Fig. 3(a) in the main text). As shown in Fig. S7(b), a clear THG signal is observed when the pump laser is tuned to the fundamental mode (right). In contrast, no detectable THG signal is observed when the same pump power is coupled into the second-order cavity mode (left). This behavior can be explained by the substantially lower *Q*/*V* ratio of the second-order mode. The experimentally measured *Q* factors of the fundamental (first-order) and second-order cavity modes for the *d5* coupling configuration are 28,000 and 2,500, respectively, while the corresponding mode volumes obtained from numerical simulations are $0.96(\lambda/n)^3$ and $3.8(\lambda/n)^3$. The corresponding mode profiles are shown in the inset of Fig. S7(a). Consequently, the *Q*/*V* ratio of the second-order mode is approximately 44 times smaller than that of the fundamental mode, making efficient THG generation difficult. The electric-field components of the fundamental mode are also shown in Fig. S7(c). In the simplified surface-nonlinearity model used in the main text, the SHG contribution is represented by the electric-field component normal to the photonic-crystal air-hole surfaces. Thus $E_{\perp}$ represents the surface-sensitive field component, whereas the total field amplitude $E$ is used for the bulk THG contribution. These results provide further experimental evidence that the optimized cavity achieves an exceptionally large $Q/V$ ratio. Figure S7(d) summarizes the THG measurements obtained from four different devices with cavity–waveguide separations of d4 and d5. As shown in the figure, THG was successfully observed in multiple devices, demonstrating the reproducibility and robustness of the optimized cavity design.

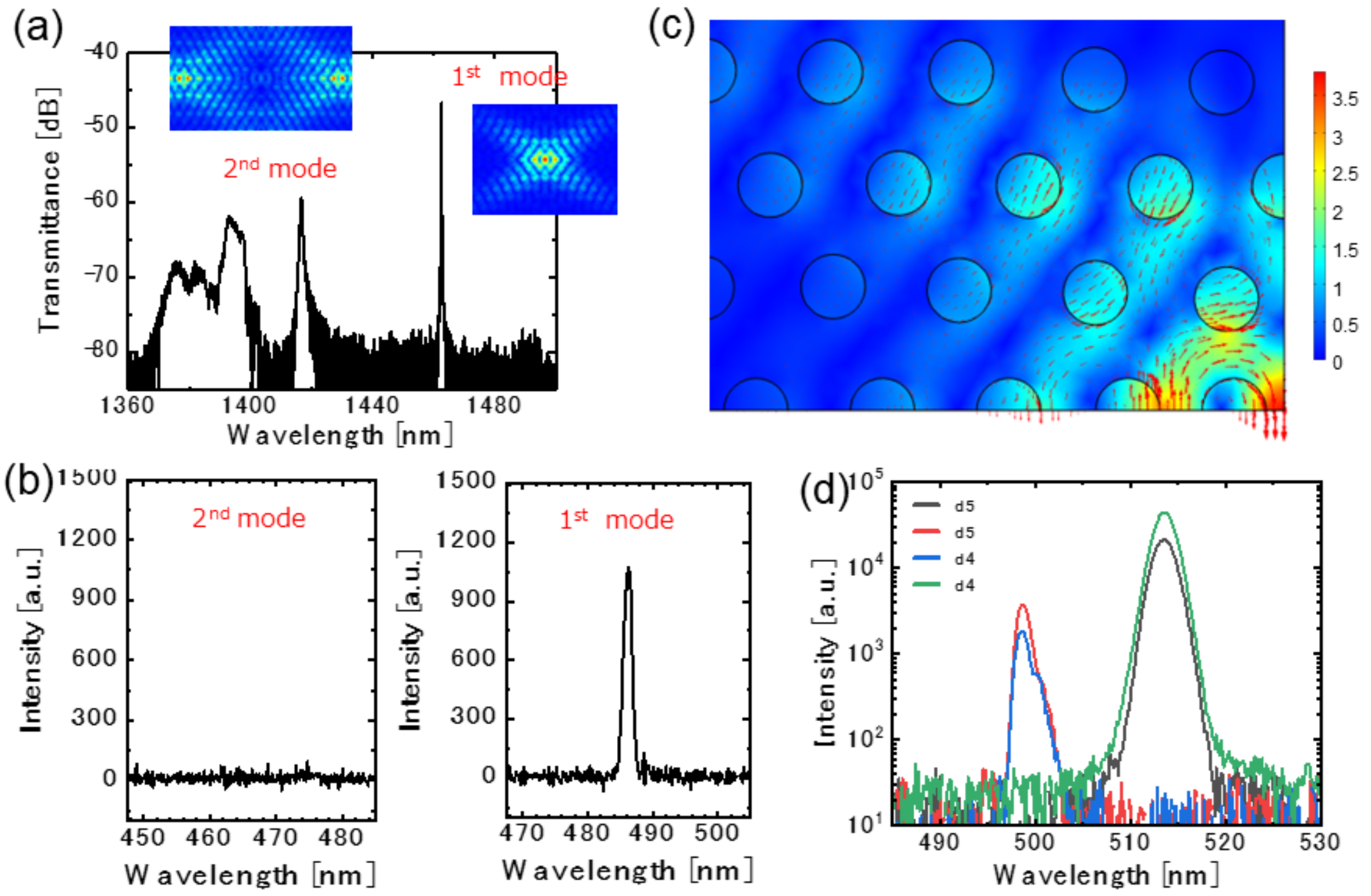

Fig. S7. **Characterization of third-harmonic generation (THG).**
(a) Transmission spectrum of the SiN photonic crystal cavity. The electric-field mode profiles of the first- and second-order cavity modes are shown as insets. (b) THG spectra obtained by resonantly exciting the first- and second-order cavity modes. (c) Electric-field profile of the first-order cavity mode, together with arrows indicating the electric-field components. (d) THG spectra obtained by resonantly exciting the first-order cavity mode in four different samples.